\documentclass[preprint,12pt]{elsarticle}

\usepackage{amssymb}
\usepackage{bm}
\usepackage{amsmath}
\usepackage[utf8x]{inputenc} 

\journal{Physics Letters A}

\begin{document}

\begin{frontmatter}

\title{Spectroscopy of the Dirac oscillator perturbed by a surface delta
potential}

\author{J.\ Mun\'{a}rriz and F Dom\'{\i}nguez-Adame}

\address{GISC. Departamento de F\'{\i}sica de Materiales, Universidad
Complutense, E-28040 Madrid, Spain}

\author{R.\ P.\ A.\ Lima}

\address{Instituto de F\'{\i}sica, Universidade Federal de Alagoas,
Macei\'{o}, AL 57072-970, Brazil}

\begin{abstract}

We study theoretically the level shift of the Dirac oscillator perturbed by any
sharply peaked  potential approaching a surface delta potential. A Green
function method is used to obtain closed expressions for all partial waves and
parities. 

\end{abstract}


\end{frontmatter}


According to their behavior under Lorentz transformations, the interaction
potentials for the Dirac equation are either vector, scalar or pseudoscalar. 
Here the term vector means the time component of a Lorentz vector, while a
scalar potential is equivalent to a dependence of the mass upon position.
Relativistic wave equations with vector and scalar linearly rising potentials
have been widely used to investigate the confinement of particles in nuclear and
hadron physics. When the linear potential is vectorlike, there exist no bound
states and only tunneling solutions arise~\cite{Glasser84,Adame92a}. Therefore,
vector linear potentials cannot confine particles, being another fine example of
the Klein tunneling~\cite{Galic88}. On the contrary, scalar linear potentials
can bind relativistic particles and give rise to confinement since the mass of
the particle increases without bound when the separation from the center gets
larger~\cite{Adame92a,Critchfield76,Ram87}. Similar conclusions can be drawn for
vector and scalar parabolic potentials~\cite{Toyama99}.

In spite of the fact that vector linear and quadratic potentials cannot bind
particles, it is possible to extend the quantum harmonic oscillator to the
relativistic domain~\cite{Toyama99}. In this context, Moshinsky and Szczepaniak
introduced a new type of interaction in an attempt to describe a
relativistic oscillator by means of a Dirac equation linear in both momenta and
coordinates~\cite{Moshinsky89}. The authors gave the name of Dirac oscillator to
this model, whose nonrelativistic limit leads to a standard harmonic oscillator
equation with spin-orbit term. The complete energy spectrum and the
electromagnetic potential associated to the Dirac oscillator were found by
Ben\'{\i}tez \emph{et al.}~\cite{Benitez90}. The spectrum presents degeneracies which
are explained by a symmetry Lie algebra~\cite{Quesne90}. Regardless the
intensity of the coupling, this interaction leaves the vacuum unchanged and the
Klein tunneling is avoided~\cite{Benitez90,Adame92b}. Therefore, in contrast to
the vector linear potential, the resulting states are truly bound and the
Dirac oscillator is a good candidate to explain the observed confinement of
quarks~\cite{Moreno89}. This interaction has also been considered in the
two-body Dirac equation, presenting interesting features which are not shared
with the one-body oscillator~\cite{Adame91a}. Recently, a photonic realization
of the Dirac oscillator based on light propagation in engineered fiber Bragg
gratings has been discussed by Longhi~\cite{Longhi10}. Therefore, the interest
of the Dirac oscillator is rather general and well beyond high energy physics.

The aim of this paper is to study the spectroscopy of the Dirac oscillator in a
($3+1$)-dimensional space perturbed by a surface delta potential, solving the
associated Lippmann-Schwinger equation. Therefore, we present a generalization
of the results obtained in Ref.~\cite{Adame91b}, in the sense that we do not
restrict ourselves to ($1+1$)-dimensional space. In addition, in
Ref.~\cite{Adame91b} the perturbation was described by a nonlocal separable
potential. The $\delta$-function limit of a nonlocal potential in the Dirac
equation is mathematically well defined~\cite{Calkin88} and the corresponding
Lippmann-Schwinger equation is valid even in this limit. However, as discussed
in Ref.~\cite{Adame90}, there exist some ambiguities in defining the surface
delta potential (local) that require a careful analysis of the
Lippmann-Schwinger equation, as we show below.


We arrive at the Dirac oscillator equation by the non-minimal substitution ${\bm
p}\to {\bm p} -im\omega \beta{\bm r}$, where $m$ is the mass of the particle,
$\omega$ is the oscillator frequency and $\beta$ is the usual Dirac matrix
defined below. From the above considerations, the Dirac Hamiltonian may be
written in the standard notation as (in units with $\hbar=1$ and $c=1$)
\begin{equation}
\mathcal{H}={\bm\alpha} \cdot ({\bm p}-im\omega \beta {\bm r})+\beta m+V_R(r) 
\equiv 
\mathcal{H}_0 + V_{R}(r)\ .
\label{Dirac_Hamiltonian}
\end{equation}
We choose the representation 
$$
{\bm \alpha}=
\left(
\begin{array}{cc}
0           & {\bm \sigma}\\
{\bm \sigma} & 0
\end{array}
\right)\ ,
\qquad
{\bm \beta}=
\left(
\begin{array}{cc}
I_2 & 0\\
0   & -I_2
\end{array}
\right)\ ,
$$
where ${\bm \sigma}=(\sigma_x,\sigma_y,\sigma_z)$ contains the Pauli matrices
and $I_2$ is the $2\times 2$ unity matrix.  

We are interested in the energy spectrum of the Dirac oscillator, whose
Hamiltonian is given by $\mathcal{H}_0$ in~(\ref{Dirac_Hamiltonian}), perturbed
by a potential $V_R(r)$ approaching the $\delta$-shell limit with radius $R$.
However, as mentioned above, the resulting equation is ambiguous if one takes
the limit $V_R(r)\to \lambda \delta(r-R)$ from the outset. The origin of the
ambiguity is the following. Since the Dirac equation is linear in momentum, the
wavefunction itself must be discontinuous at $r=R$ to account for the singular
term $V_R(r)\to \lambda \delta(r-R)$. However, the product of a discontinuous 
function and the
$\delta$-function is mathematically ill defined. This ambiguity can be avoided
by solving the corresponding Dirac equation for any arbitrary sharply peaked at
$r=R$ function, $R$ being the radius of the shell, and then take the
$\delta$-function limit with the constraint
\begin{equation}
\int_{R-\Delta R}^{R+\Delta R} V_R(r)\,dr = \lambda\ ,
\qquad
\Delta R \to 0 \ ,
\label{condition}
\end{equation}
where $\lambda$ is the dimensionless coupling constant.

For the moment, we only assume that the potential $V_R(r)$ is spherically
symmetric. The eigenfunctions of definite parity and total angular momentum
$(J^2,J_z)$ are written in the form
\begin{equation}
{\bm \Psi}({\bm r})=\frac{1}{r}
\left(
\begin{array}{c}
if(r)\\
 g(r) {\bm \sigma}\cdot {\bm r}/r
\end{array}
\right){\bm \Phi}_{jm_j}^{l}\ ,
\label{full-spinor}
\end{equation}
where $\bm \Phi_{jm_j}^{l}$ are the normalized two-components eigenfunctions 
of $J^2$, $J_z$,
$L^2$ and $S^2$~\cite{Bjorken64}. Using~(\ref{Dirac_Hamiltonian})
and~(\ref{full-spinor}), the Dirac equation leads to
\begin{equation}
\frac{d\phantom{r}}{dr}\,{\bm \phi}(r)=\left[ \sigma_z\gamma(r) 
-\sigma_x m+i\sigma_y\Big(E-V_R(r)\Big)\right]{\bm \phi}(r) \ ,
\label{radial_equation_1}
\end{equation}
where the upper and lower components of the radial spinor ${\bm \phi}(r)$ are
$f(r)$ and $g(r)$, respectively. Here $\kappa=\mp (j+1/2)$ for $l=j\pm 1/2$
and we have defined $\gamma(r)=\kappa/r+m\omega r$ for brevity.
Equation~(\ref{radial_equation_1}) is solved by a Newmann solution as
follows~\cite{McKellar87}
$$
{\bm \phi}(r)=\widehat{P}\,\exp\Bigg\{
\int_{r_0}^{r} \!dr^{\prime} \left[ \sigma_z \gamma(r^{\prime})
-\sigma_x m+i\sigma_y\Big(E-V_R(r^{\prime})\Big)\right]\Bigg\}
\,{\bm \phi}(r_0) \ ,
$$
where $\widehat{P}$ is the ordering operator. Setting $r=R+\Delta R$ and
$r_0=R-\Delta R$, taking the limit $\Delta R\to 0$ and using the
constraint~(\ref{condition}) we finally obtain the  following boundary condition
\begin{equation}
{\bm \phi}(R+\Delta R)=\exp\left(-i\lambda \sigma_y\right)
{\bm \phi}(R-\Delta R)\ ,
\label{boundary_condition}
\end{equation}
which becomes independent of how the $\delta$-function limit is taken and thus
we avoid any ambiguity defining the relativistic surface $\delta$ potential.

Once the correct boundary condition at the singularity has been obtained, we
proceed to find the energy spectrum of the perturbed Dirac oscillator. To this
end, we write the Lippmann-Schwinger solution of the radial Dirac
equation~(\ref{radial_equation_1})
\begin{equation}
{\bm \phi}(r)=-\int_{R-\Delta R}^{R+\Delta R}
\mathcal{G}(r,r^{\prime};E) V_R(r^{\prime}){\bm \phi}(r^{\prime})dr^{\prime}\ ,
\label{solution_1}
\end{equation}
where the Green function for the unperturbed problem is a $2\times 2$ matrix
satisfying the inhomogeneous differential equation
\begin{equation}
\left[-i\sigma_y\frac{\partial\phantom{r}}{\partial r}+\sigma_x\gamma(r)+\sigma_z m-E\right]
\mathcal{G}(r,r^{\prime};E)=I_2\delta(r-r^{\prime})\ .
\label{green_1}
\end{equation}
The Green function exhibits
a jump discontinuity at the line $r=r^{\prime}$. The value of the jump can be
obtained by integration of~(\ref{green_1}) in the vicinity of this line. The
result is
\begin{equation}
\mathcal{G} (r+\Delta R,r;E)-\mathcal{G}(r-\Delta R,r;E)=i\sigma_y \ .
\label{jump}
\end{equation}

The product $V_R(r){\bm \phi}(r)$ in the integral~(\ref{solution_1}) is  not
well defined if one takes the limit $V_R(r)\to \lambda \delta(r-R)$, as we
already discussed. Thus, we consider the same limiting  procedure discussed 
previously and solve~(\ref{solution_1}) for any arbitrary sharply peaked at
$r=R$ function and then take the $\delta$-function limit. Using the
radial Dirac equation~(\ref{radial_equation_1}) one finds that the
integral equation~(\ref{solution_1}) leads to (see  Ref.~\cite{Adame92c} for
details)
\begin{equation}
{\bm \phi}(r)=i\mathcal{G} (r,R;E)\sigma_y
\Big[{\bm \phi}(R+\Delta R)-{\bm \phi}(R-\Delta R)\Big]\ .
\label{eigenfunction_1}
\end{equation}
Hence we have obtained a closed expression for the perturbed eigenfunctions. The
energy levels can be obtained by setting $r=R+\Delta R$ 
in~(\ref{eigenfunction_1}) and using the boundary
condition~(\ref{boundary_condition})
\begin{equation}
\det\Big[I_2-i\mathcal{G} (R+\Delta R,R;E)
\sigma_y\left(e^{i\sigma_y\lambda} - I_2\right)\Big]=0\ .
\label{levels}
\end{equation}

Therefore, the energy spectrum of the perturbed Dirac oscillator can be obtained
provided the Green function for the unperturbed problem is known. The Green
function of the Dirac oscillator can be cast in the form
$$
{\mathcal G}(r,r^{\prime};E)=\left(
                          \begin{array}{cc}
                             G_{++}(r,r^{\prime};E) & G_{+-}(r,r^{\prime};E)\\
                             G_{-+}(r,r^{\prime};E) & G_{--}(r,r^{\prime};E)
                          \end{array}
                     \right)
$$
and it was calculated in Ref.~\cite{Alhaidari04}.
For the sake of brevity we only quote here the final expressions of the
diagonal terms
\begin{subequations}
\begin{equation}
\label{diagGreen}
G_{\pm \pm} = \frac{E \pm m}{m \omega}
  \frac
    {\Gamma \left(\nu_\pm -\mu_\pm + \frac{1}{2}\right)}
    {\Gamma \left(2 \nu _\pm +1\right) \sqrt{r r'}}\,
  M_{\mu _\pm,\nu _\pm}\left(m \omega r_<^2 \right) 
    W_{\mu _\pm,\nu _\pm}\left(m \omega r_>^2 \right)\ ,
\end{equation}
where $M_{\mu,\nu}$ and $W_{\mu,\nu}$ are the Whittaker functions, $r_<=\min(r,r')$ and
$r_>=\max(r,r')$.  The parameters $\mu_\pm$ and $\nu_\pm$ are defined as
\begin{align}
  \mu_\pm &= \frac{1}{4} \left( \frac{E^2-m^2}{m\omega}
    - 2\kappa \pm 1 \right) \ , &
  \nu_\pm &= \frac{1}{2} \left| \kappa \pm \frac{1}{2}\right|\ .
\end{align}
The off-diagonal terms are calculated from the following expressions
\begin{equation}
\label{nonDiagGreen}
G_{\pm \mp} = \frac{1}{E \mp m} \left[ \mp \frac{\partial}{\partial r} +
  \frac \kappa r + m\omega r \right] G_{\mp \mp}\ .
\end{equation}
\label{allGreen}
\end{subequations}

\begin{figure}[ht]
\centerline{\includegraphics[width=0.68\columnwidth]{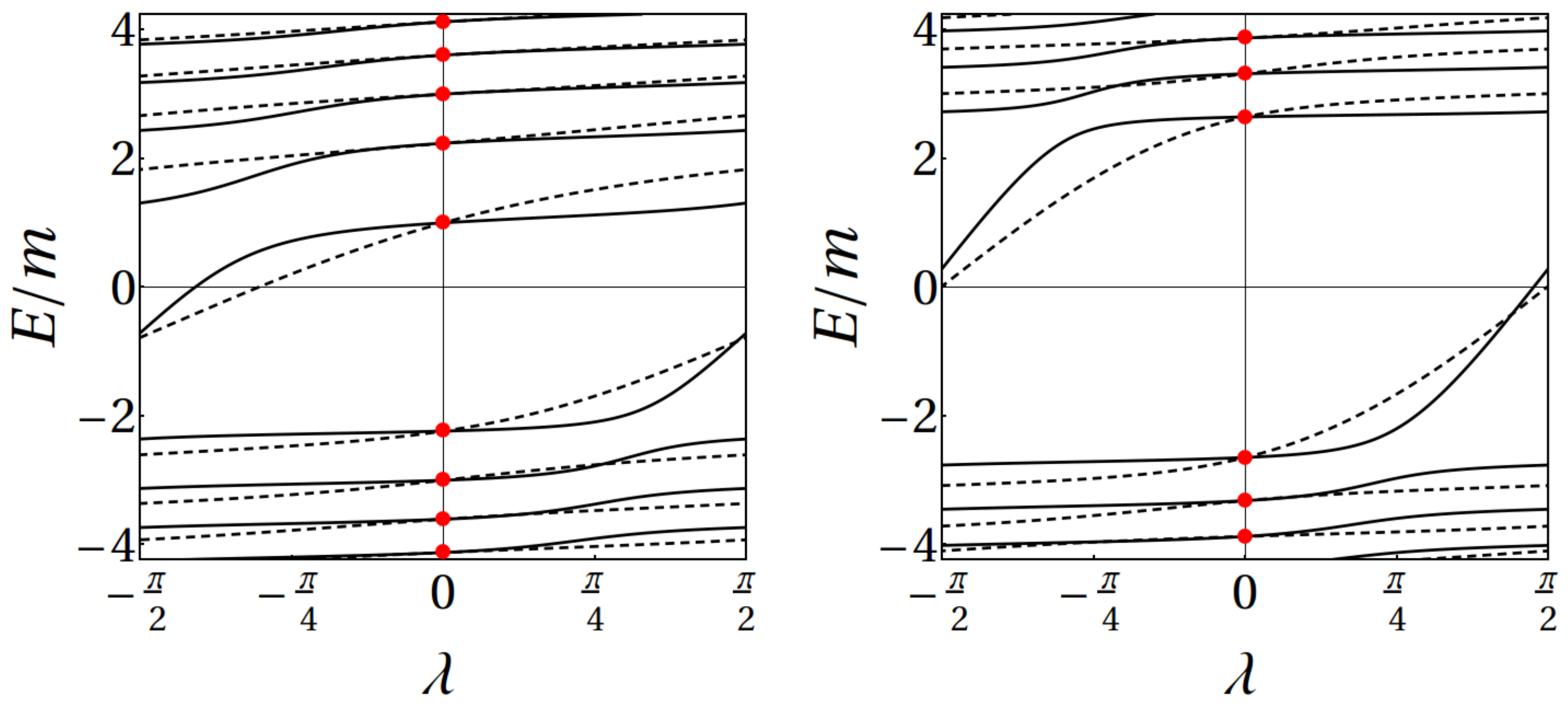}}
\caption{Energy levels of a Dirac oscillator perturbed by a surface delta
potential as a  function of the coupling constant $\lambda$ for $\omega=m$.
Left and right panels correspond to $\kappa=-1$ and $\kappa=1$, respectively.
Solid (dashed) lines correspond to the results for $R=0.3/m$ ($R=1/m$).
The solutions of the unperturbed Dirac oscillator are marked with red points.}
\label{fig1}
\end{figure}

Real solutions of~(\ref{levels}) with the Green function given
by~(\ref{allGreen}) yield the energy levels of the
perturbed Dirac oscillator.  Due to the boundary
condition~(\ref{boundary_condition}), the energy levels of the perturbed Dirac
oscillator satisfy the property $E(m,\kappa,R,\lambda) = 
E(m,\kappa,R,\lambda + \ell \pi)$, $\ell$ being an integer. Since
the energy levels are $\pi$-periodic functions of the coupling constant, we can restrict
ourselves to the range $-\pi/2 < \lambda \leq \pi/2$ hereafter.
Figure~\ref{fig1} shows these levels as a function
of the coupling constant for the case $\omega=m$. The energy levels of the 
perturbed Dirac oscillator are shifted upwards on increasing the coupling
constant from $-\pi/2$ to $\pi/2$. Notice that the levels of the perturbed Dirac
oscillator cross those of the unperturbed oscillator only when $\lambda=n\pi$,
$n$ being an integer. According to Eq.~(\ref{boundary_condition}), in this
case ${\bm \phi}(R+\Delta R)=(-1)^n{\bm \phi}(R-\Delta R)$ and the
surface $\delta$ potential is actually transparent.

In the limit $\lambda \rightarrow 0$, the eigenvalues of the unperturbed
system are recovered~\cite{Szmytkowski01}.  They correspond to the poles
of $\mathcal G$, namely the set of values
$E_0(m,\kappa)$ for which $(E_0 \pm m) \Gamma(\nu_\pm - \mu_\pm +1/2)
\rightarrow \infty$
\begin{equation}
  E_0^2(m,\kappa) - m^2 =
    4m \omega \left[ n + \theta(\kappa) \left( 2\kappa + 1 \right) \right]\ ,
   \quad E_0\neq -m\ ,
\end{equation}
where $\theta$ is the Heaviside step function and $n$ is a nonnegative integer.

\begin{figure}[ht]
\centerline{\includegraphics[width=0.35\columnwidth]{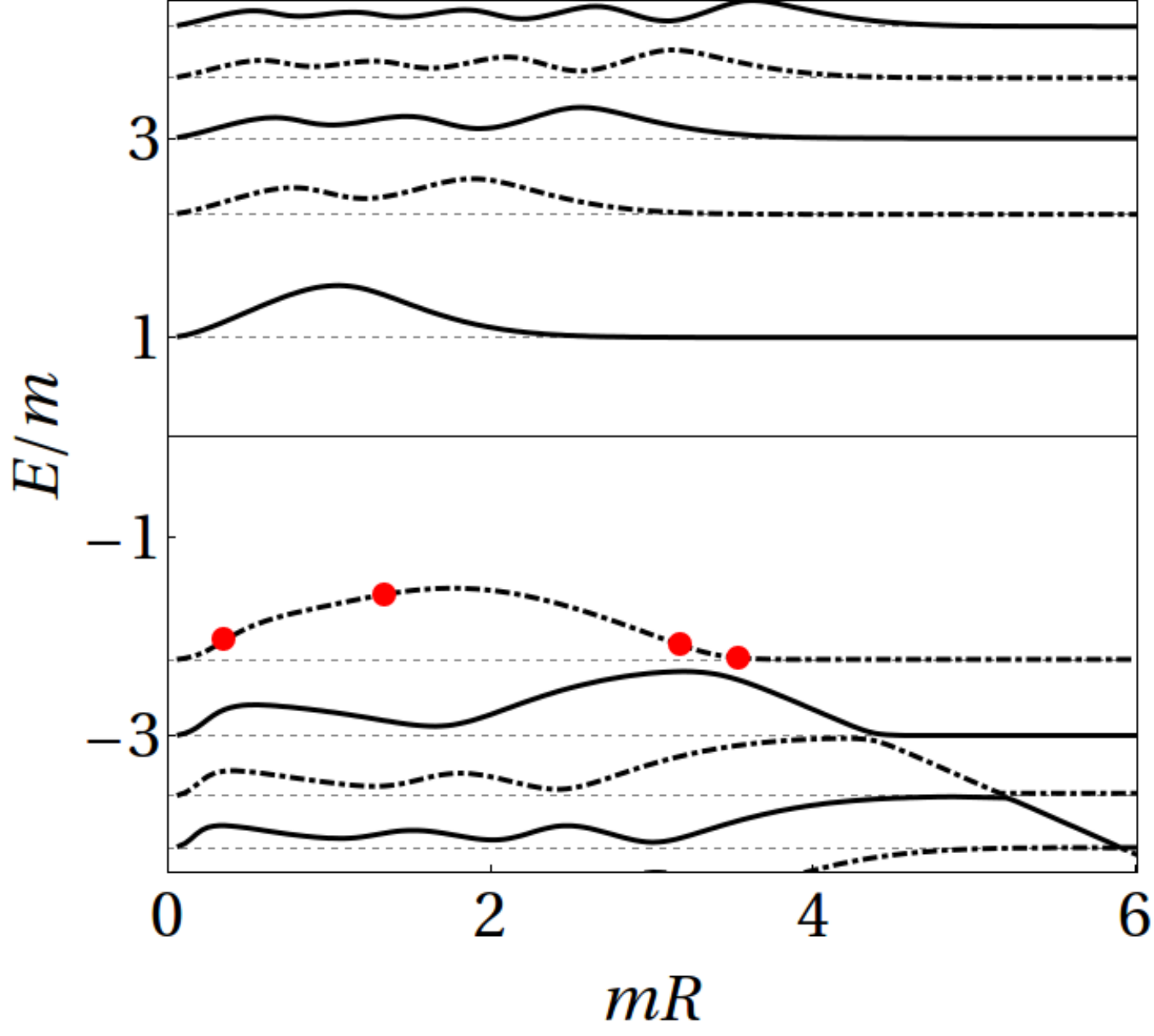}}
\caption{Shift of the energy levels as a function of the 
radius $R$  for $\omega=m$, $\kappa=-1$ and $\lambda=\pi/4$.  
The levels of the unperturbed Dirac
oscillator are plotted with dashed gray lines.  The wave functions corresponding
to the four red points are shown in Fig.~\ref{fig3} The wave functions at points marked
$a-d$ are shown in figure~\ref{fig3} below.}
\label{fig2}
\end{figure}

Due to the confining properties of the Dirac oscillator, the spatial extent of the
eigenstates increases with the absolute value of the energy, $\left|E\right|$. 
If this spatial extent is much smaller than the radius of the surface $\delta$ potential, 
the net effect of the perturbation on the eigenfunction is small, and therefore the energy
does not change noticeably. This trend is clearly observed in Fig.~\ref{fig2}, where the
energy levels are plotted as a function of the radius $R$
for $\omega=m$, $\kappa=-1$ and $\lambda=\pi/4$. 
When the dimensionless coupling constant $\lambda$ is large, the energy levels display anticrossings,
as seen in the figure. Moreover, each energy level approaches the same
level of the unperturbed Dirac oscillator in the two limiting situations, $mR\ll 1$ and $mR\gg 1$.

\begin{figure}[ht]
\centerline{\includegraphics[width=0.5\columnwidth]{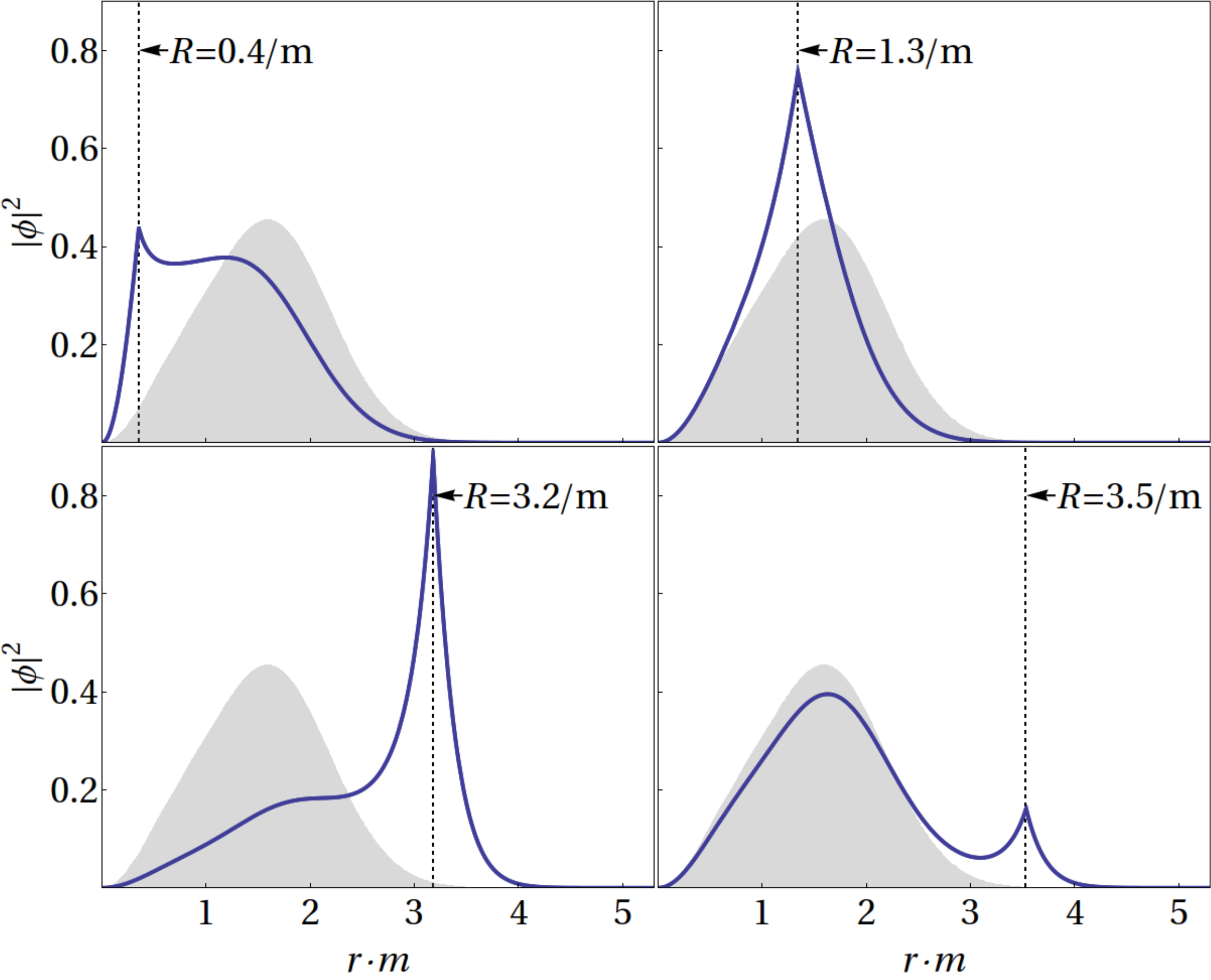}}
\caption{Plots of $|\bm \phi(r)|^2$ as a function of the  dimensionless radial
coordinate $z=r\cdot m$ for the four different values of the delta-shell radius
$R$ indicated in figure~\ref{fig2} by red circles, when $\omega=m$, $\kappa=-1$
and $\lambda=\pi/4$. The corresponding unperturbed eigenstate is plotted using 
gray solid regions.}
\label{fig3}
\end{figure}

Finally, the change of the probability density $|\bm \phi(r)|^2$ due to 
the surface $\delta$ potential is shown in Fig.~\ref{fig3}.  The set of parameters chosen
correspond to the four red circles in Fig.~\ref{fig2}, {\it i.e.}, all of them
are taken from the same original unperturbed eigenstate.
As $R$ is increased, the eigenstate transits between two
loosely perturbed states -- upper left and lower right panels. In between, states
with probability density strongly peaked around $r=R$ are found.

In summary, we have calculated the shift of the energy levels of the Dirac oscillator 
perturbed by a surface $\delta$ potential using a Green function technique. The method
is valid for any sharply peaked potential approaching the $\delta$-function 
and consequently it is free of the ambiguities
appearing in defining relativistic $\delta$-interactions~\cite{Adame90}. 
Remarkably, the energy spectrum is a $\pi$-periodic function of the coupling constant
$\lambda$, a situation not found in the ($1+1$)-dimensional Dirac oscillator
perturbed by a nonlocal $\delta$ potential~\cite{Adame91b}.

\bigskip

Work in Madrid was supported by MICINN (project MAT2010-17180). R. P. A. Lima
would like to thank CAPES via project PPCP-Mercosul,  CNPq, and FINEP (Brazilian
Research Agencies) as well as FAPEAL (Alagoas State Research Agency) for partial financial support.

\end{document}